\documentclass[sigconf]{acmart}

\acmConference[ASE '26]{41th IEEE/ACM International Conference on Automated
  Software Engineering}{October 2026}{Munich, Germany}

\setcopyright{none}
\renewcommand\footnotetextcopyrightpermission[1]{}

\usepackage{booktabs}
\usepackage{graphicx}
\usepackage{pifont}
\usepackage{pgfplots}
\pgfplotsset{compat=1.18}
\usepackage{tikz}
\usepackage{makecell}
\usepackage{array}
\usepackage{tabularx}

\title[JavaVulBench: A Realistic Java Vulnerability Benchmark]{JavaVulBench: A Java Vulnerability Benchmark with Realistic
       Splits, a Unified Multi-Backend Harness, and a Leakage-Aware
       Evaluation Mode}

\author{Norbert Sándor Szolnoki}
\authornote{Both authors contributed equally to this research.}
\email{szolnoki@inf.u-szeged.hu}
\orcid{0009-0000-6351-9000}
\author{Gábor Antal}
\authornotemark[1]
\email{antal@inf.u-szeged.hu}
\affiliation{%
  \institution{University Of Szeged}
  \city{Szeged}
  \country{Hungary}
}

\begin{document}

\begin{abstract}
We release \textsc{JavaVulBench}, a benchmark dataset and evaluation
harness for Java vulnerability detection. The dataset contains
$\sim$30{,}600 Java methods spanning 1{,}740 CVEs and 700+ projects,
labelled at both method and line granularity, with per-CVE publication
dates and five realistic split strategies: random, project-disjoint,
temporal, deduplicated, and unseen CWE-family. The harness provides a
single \texttt{LlmPrediction} schema across three backend families
(encoder classifiers, local generative models served by Ollama, and
API-served LLMs routed through OpenRouter) so that twelve reference
detectors CodeBERT, GraphCodeBERT, UniXcoder, DeepSeek-Coder-1.3B,
and eight API/open-weight LLMs (GPT-4o, GPT-4.1-mini, Claude Sonnet~4,
DeepSeek-v3, DeepSeek-Coder-v2, Qwen-2.5-Coder-14B/7B,
CodeLlama-13B) are evaluated under identical conditions from a single
command. A pre-training contamination audit is shipped alongside every
model so users can separate genuinely unseen test CVEs from potentially
memorised ones. Data, code, and fine-tuned checkpoints are archived on
Zenodo~\cite{javavulbench2026} and short
demonstration video is available on YouTube (https://www.youtube.com/watch?v=nMTX\_hqkuoM)\footnote{\url{https://www.youtube.com/watch?v=nMTX_hqkuoM}}.
\end{abstract}

\maketitle

\section{Introduction}
\label{sec:intro}

Deep-learning-based vulnerability detection has been benchmarked
predominantly on C/C++ corpora such as Big-Vul~\cite{bigvul2020} and
Devign~\cite{devign2019}. Java is a widely-deployed enterprise
language and the OWASP Top-10 categories~\cite{owasptop10}
(injection, deserialisation, SSRF, path traversal,
authentication/authorisation) commonly appear in disclosed Java CVEs,
yet
Java-focused evaluation infrastructure is comparatively thin. When Java
is evaluated at all it is typically through a random split of the
multilingual CVEfixes corpus~\cite{cvefixes2021}, with no
project-disjoint or temporal partition. Chakraborty et
al.~\cite{chakraborty2024} showed on C/C++ that this single
methodological choice can inflate reported F1 by 15-40 points, because
near-duplicate methods drawn from the same project routinely leak
between train and test.

To the best of our knowledge, existing Java-focused benchmarks -
Vul4J~\cite{vul4j2022}, VJBench~\cite{vjbench2023},
CWE-Bench-Java (released with the IRIS system)~\cite{cwebenchjava2024},
and the Java slice of CVEfixes~\cite{cvefixes2021} - do not simultaneously ship multiple
realistic split strategies, explicit non-vulnerable samples,
line-level labels, a unified multi-backend evaluation harness, and a
per-model pre-training contamination audit. Table~\ref{tab:related}
summarises this comparison. \textsc{JavaVulBench} is built to fill that
combination.

\textbf{Intended users.} Vulnerability-detection researchers who need a
Java benchmark with \emph{controlled} leakage, and tool builders who
want a drop-in harness that evaluates their detector against existing
baselines under identical conditions.

\textbf{Contributions.} We release \textsc{JavaVulBench}, consisting of:
(i) a CVE-grounded corpus of $\sim$30{,}600 Java methods spanning
1{,}740 CVEs and 700+ projects, labelled at both method and line
granularity, with per-sample publication dates and five realistic
split strategies (random, project-disjoint, temporal, deduplicated at
Jaccard 0.8, unseen CWE-family);
(ii) a unified evaluation harness with a single \texttt{LlmPrediction}
schema spanning three backend families - HuggingFace encoder
classifiers (CodeBERT~\cite{codebert2020},
GraphCodeBERT~\cite{graphcodebert2021}, UniXcoder~\cite{unixcoder2022}),
Ollama~\cite{ollama}-served DeepSeek-Coder-1.3B
\cite{deepseekcoder2024}, and eight OpenRouter~\cite{openrouter}-served
API LLMs (GPT-4o~\cite{gpt4o2024}, GPT-4.1-mini, Claude Sonnet~4,
DeepSeek-v3~\cite{deepseekv3}, DeepSeek-Coder-v2~\cite{deepseekcoderv2},
Qwen-2.5-Coder-14B/7B~\cite{qwen25coder},
CodeLlama-13B~\cite{codellama2023}) - driven from a single command;
(iii) a \emph{pre-training contamination audit} that partitions every
test split per model into \emph{risky} (CVEs before the model's cutoff)
and \emph{clean} (after) subsets with leakage-aware metric reporting;
and (iv) a Dockerised, one-command reproduction pipeline with
fine-tuned checkpoints archived on Zenodo~\cite{javavulbench2026}.

Our headline empirical finding (Section~\ref{sec:vs1}) is that switching
from a random to a project-disjoint split on Java reproduces the 15-40
F1-point gap Chakraborty et al.~\cite{chakraborty2024} observed on
C/C++. The harness and contamination audit are the infrastructure that
make this and similar studies reproducible.

\section{Dataset}
\label{sec:dataset}

\textbf{Mining.} CVE records are merged from three sources: the NIST
National Vulnerability Database~\cite{nvd}, the GitHub Security
Advisory database~\cite{ghsa}, and CVEfixes~\cite{cvefixes2021}. Each
record resolves to one or more fix commits; pre- and post-fix Java
source snapshots are pulled through the GitHub API and cached. A
tree-sitter~\cite{treesitter} grammar extracts method bodies with
method-, class-, and file-level provenance preserved. When a commit's
source files are unavailable the harness falls back to diff-level
parsing. A noise filter then removes test-only, formatting-only and
refactoring-only commits, which we found routinely inflate earlier
corpora.

\textbf{Labelling.} Each vulnerable method is paired with its patched
counterpart under a shared \texttt{pair\_id}. Negatives are drawn from
three strategy per vulnerable method: same file (3:1), same project
(2:1), and a random project (1:1). Every sample carries the CVE's
publication date, the CWE tag, and the set of vulnerable line numbers
from the fix commit. The final corpus contains 30{,}637 methods
(5{,}281 vulnerable, 4{,}964 patched, 20{,}392 negatives) across
700+ projects and 225 CWE categories from 2015-2025.

\textbf{Splits.} We ship five partition strategies as ready-to-use
JSONL. \emph{Random} is the optimistic reference; \emph{project}
guarantees no project overlap between train and test; \emph{time}
trains on CVEs published before 2023 and tests on those from
2023-07 onwards; \emph{deduplicated} applies Jaccard-0.8
near-duplicate removal before the split; and \emph{CWE-family} holds
out entire CWE families (e.g.\ all CWE-79 XSS samples) so models are
forced to generalise across vulnerability classes. This five-way
release makes the split strategy itself a first-class, reportable
property of any benchmark number.

\textbf{Datasheet.} The corpus ships with a datasheet in the
Gebru et al.~\cite{datasheets2021} template covering motivation,
composition, collection, pre-processing, intended uses and
maintenance, including known bias risks (English-only commit
messages; Apache/GitHub-heavy ecosystem).

\section{Evaluation Harness}
\label{sec:harness}

The evaluation harness exposes a unified
\texttt{LlmPrediction} interface across all backends, enabling the
same evaluation, plotting, and statistical-analysis pipeline to be
reused unchanged. Integrating a new detector requires implementing a
single \texttt{predict} interface returning the predicted label,
confidence, optional CWE category, and optional vulnerable lines.

\textbf{Backends.} The framework supports three backend families:
\emph{(i)} HuggingFace Transformers for encoder-style classifiers,
including CodeBERT~\cite{codebert2020},
GraphCodeBERT~\cite{graphcodebert2021},
UniXcoder~\cite{unixcoder2022}, and compatible checkpoints such as
PDBERT~\cite{pdbert2024};
\emph{(ii)} Ollama~\cite{ollama} for local generative inference,
including DeepSeek-Coder-1.3B~\cite{deepseekcoder2024} in zero-shot and
LoRA-fine-tuned~\cite{lora2022} configurations; and
\emph{(iii)} OpenRouter~\cite{openrouter} for API-based LLM evaluation,
covering GPT-4o~\cite{gpt4o2024}, GPT-4.1-mini, Claude Sonnet~4,
DeepSeek-v3~\cite{deepseekv3},
DeepSeek-Coder-v2~\cite{deepseekcoderv2},
Qwen-2.5-Coder~\cite{qwen25coder}, and
CodeLlama-13B~\cite{codellama2023}.

\textbf{Prompt strategies.} Generative backends support zero-shot,
few-shot, and retrieval-augmented generation (RAG)~\cite{rag2020}
evaluation. The RAG pipeline combines sentence-transformer
embeddings~\cite{sbert2019} indexed with
FAISS~\cite{faiss2021}, BM25 lexical
retrieval~\cite{bm25}, and Reciprocal Rank
Fusion~\cite{rrf2009}. The knowledge base is intentionally small and
Java-specific, containing 18 MITRE CWE descriptions mapped to common
Java APIs/framework idioms (e.g.,
\texttt{ObjectInputStream.readObject},
\texttt{DocumentBuilderFactory}, JDBC query concatenation) and 10
OWASP/CERT-derived API-misuse patterns paired with vulnerable/secure
Java examples. The collection focuses on Java-relevant OWASP Top 10
categories and excludes non-Java or overly abstract CWEs.

\textbf{Cost control.} API-based evaluations use a fixed stratified
200-sample subset per split (seed 42), allowing direct comparison
across models and prompt strategies while keeping large-scale API
evaluation financially tractable. All API calls are routed through
OpenRouter behind a single interface.

\textbf{Metrics.} The harness reports detection metrics
(precision, recall, F1, MCC~\cite{matthews1975}, PR-AUC,
false-positive rate), calibration metrics (ECE~\cite{guo2017},
Brier score~\cite{brier1950}), localization metrics (top-$k$
accuracy, IoU, MAP@$k$), and CWE metrics. MCC summarises all
confusion-matrix cells under imbalance; PR-AUC measures
threshold-independent ranking; ECE and Brier assess confidence
calibration, where lower is better.
\section{Validation Studies}
\label{sec:validation}

To demonstrate that the artefact functions as claimed we run three
validation studies, each reproducible with a single script invocation
and writing its tables as CSV files directly into \texttt{results/}. The
central claim we validate is that \emph{on Java, split strategy is a
first-class property of any reported number}; the harness
(Section~\ref{sec:vs2}) and the leakage-aware mode
(Section~\ref{sec:vs4}) are the enabling infrastructure around that
claim.

\subsection{VS1: Split strategy is a first-class property}
\label{sec:vs1}

Table~\ref{tab:rq1} reports detection F1 of the four fine-tuned
reference models on the full test set of each shipped split. The gap between \emph{random} and \emph{project-disjoint}
reproduces on Java the effect Chakraborty et
al.~\cite{chakraborty2024} reported on C/C++: UniXcoder drops from
F1$=0.446$ on the random split to $0.298$ on project-disjoint
($-0.148$), and GraphCodeBERT drops from $0.343$ to $0.256$
($-0.087$). CodeBERT's F1 collapses to 0 on the project-disjoint and temporal test sets: the model predicts the majority (non-vulnerable) class at the default 0.5
threshold because the held-out API surface barely overlaps with its pre-training
corpus, so recall on the positive class is zero. PR-AUC and ECE remain informative here because they expose ranking
and calibration behaviour despite the failed default threshold. The upshot for downstream users is unambiguous:
\emph{project-disjoint} should be treated as the default headline
split, with \emph{random} shipped only as an optimistic reference.

\begin{table}[t]
\centering\small
\caption{Detection F1 on the full test set of each shipped split for the
four fine-tuned reference models are distributed with the artifact. The drop
from \emph{random} to \emph{project} is the split-impact headline; it
reproduces for Java the effect Chakraborty et al.~\cite{chakraborty2024}
reported on C/C++.}
\label{tab:rq1}
\begin{tabular}{lcccc}
\toprule
Model & Random & Project & Time & CWE-family \\
\midrule
CodeBERT              & 0.422 & 0.000 & 0.000 & 0.118 \\
GraphCodeBERT         & 0.343 & 0.256 & 0.097 & 0.159 \\
UniXcoder             & 0.446 & 0.298 & 0.171 & 0.345 \\
DeepSeek-Coder-1.3B   & 0.341 & 0.062 & 0.335 & --    \\
\bottomrule
\end{tabular}
\end{table}

\subsection{VS2: Reference baselines across backend families}
\label{sec:vs2}

The unified harness lets the same split and the same
\texttt{sample\_id} set drive every backend family. Fine-tuned encoders
and DeepSeek-Coder-1.3B are evaluated on the full project-disjoint test
set (Table~\ref{tab:rq2_full}, $N=4{,}167$); these are the anchor
points new detectors should compare to. Table~\ref{tab:rq2_probe}
reports the cost-matched 200-sample probe for the eight API-served
LLMs. The two tables are not a single leaderboard: cross-family
comparisons should only use the shared probe \texttt{sample\_id} set. Claude Sonnet 4 and GPT-4o both land around
F1=0.42 on the probe, surpassing every encoder baseline. GPT-4o also shows
the best calibration of the API group (ECE=0.147), meaning that its confidence
scores are better aligned with observed correctness. DeepSeek-Coder-v2's
negative MCC indicates that, after accounting for both classes and all
confusion-matrix cells, its zero-shot decisions are below chance on Java
vulnerability detection. CodeLlama-13B has the highest PR-AUC, suggesting
that its confidence ranking is useful even though its default-threshold F1 and
calibration are weaker.

\begin{table}[t]
\centering\small
\caption{Reference detection and calibration baselines on the
project-disjoint split, full test set ($N=4167$). Higher is better
except for ECE and Brier.}
\label{tab:rq2_full}
\begin{tabular}{lccccc}
\toprule
Model & F1 & MCC & PR-AUC & ECE $\downarrow$ & Brier $\downarrow$ \\
\midrule
CodeBERT            & 0.000 & 0.000 & 0.347 & 0.086 & 0.146 \\
GraphCodeBERT       & 0.256 & 0.200 & 0.366 & 0.057 & 0.141 \\
UniXcoder           & 0.298 & 0.215 & 0.346 & 0.087 & 0.149 \\
DeepSeek-Coder-1.3B & 0.062 & 0.066 & 0.310 & 0.088 & 0.160 \\
\bottomrule
\end{tabular}
\end{table}

\begin{table}[t]
\centering\small
\caption{Cost-matched detection probe on the project-disjoint split.
All eight API-served LLMs were capped at the same deterministic
200-sample stratified subset (seed 42); running them on the full test
set would have been prohibitively expensive. Numbers are therefore
directly comparable \emph{across} API backends.}
\label{tab:rq2_probe}
\begin{tabular}{lccccc}
\toprule
Model & F1 & MCC & PR-AUC & ECE $\downarrow$ & Brier $\downarrow$ \\
\midrule
GPT-4o              & 0.419 &  0.271 & 0.438 & 0.147 & 0.204 \\
GPT-4.1-mini        & 0.400 &  0.273 & 0.465 & 0.163 & 0.178 \\
Claude Sonnet 4     & 0.416 &  0.280 & 0.349 & 0.281 & 0.293 \\
DeepSeek-v3         & 0.300 &  0.187 & 0.371 & 0.145 & 0.177 \\
DeepSeek-Coder-v2   & 0.146 & -0.108 & 0.215 & 0.312 & 0.338 \\
Qwen-2.5-Coder-14B  & 0.320 &  0.164 & 0.294 & 0.223 & 0.216 \\
Qwen-2.5-Coder-7B   & 0.247 &  0.078 & 0.195 & 0.262 & 0.255 \\
CodeLlama-13B       & 0.372 &  0.197 & 0.540 & 0.446 & 0.444 \\
\bottomrule
\end{tabular}
\end{table}

\subsection{VS3: Leakage-aware evaluation mode}
\label{sec:vs4}

The shipped contamination audit partitions each split's test set per
model into a \emph{risky} subset (CVEs published before the model's
pre-training cutoff, potentially memorised) and a \emph{clean} subset
(published after). Table~\ref{tab:leakage} reports the audit on the
\emph{time} split and the leakage-aware F1 on the 200-sample probe:
encoders with 2019--2022 cutoffs see the entire test set as clean,
while the most capable API LLMs with 2024--2025 cutoffs see up to 59\%
of the set flagged as risky, and all four strongest API models gain F1
on the clean subset --- suggesting the risky subset contains harder CVE
variants rather than easy memorisable ones.

\begin{table}[t]
\centering
\small
\caption{Per-model contamination audit on the \emph{time} test split
and leakage-aware F1 on the API-LLM probe. CVEs published $\leq$ model
cutoff are \emph{risky} (potentially memorised); the remainder is
\emph{clean}. F1$_{\text{all}}$ / F1$_{\text{clean}}$ are reported on
the 200-sample probe; encoder audit counts are on the full
$N=12{,}139$ split.}
\label{tab:leakage}

\setlength{\tabcolsep}{4pt}
\begin{tabular}{>{\raggedright\arraybackslash}p{3.2cm}crrcc}
\toprule
Model & Cutoff & Risky & Clean & F1$_{\text{all}}$ & F1$_{\text{clean}}$ \\
\midrule
\makecell[l]{CodeBERT/GraphCodeBERT/\\UniXcoder} & '19--'22 & 0 & 12{,}139 & --- & --- \\
DeepSeek-Coder-1.3B              & '23-02 & 0 & 12{,}139 & --- & --- \\
GPT-4o                           & '23-10 & 1{,}516 & 10{,}623 & 0.488 & 0.528 \\
GPT-4.1-mini                     & '24-06 & 3{,}880 & 8{,}259  & 0.296 & 0.323 \\
DeepSeek-v3                      & '24-07 & 5{,}299 & 6{,}840  & 0.302 & 0.286 \\
Claude Sonnet 4                  & '25-03 & 7{,}194 & 4{,}945  & 0.444 & 0.480 \\
DeepSeek-Coder-v2                & '24-06 & --- & --- & 0.152 & 0.213 \\
Qwen-2.5-Coder-14B               & '24-09 & --- & --- & 0.207 & 0.270 \\
Qwen-2.5-Coder-7B                & '24-09 & --- & --- & 0.138 & 0.186 \\
CodeLlama-13B                    & '23-07 & --- & --- & 0.331 & 0.331 \\
\bottomrule
\end{tabular}
\end{table}

Users reporting numbers on \textsc{JavaVulBench} are expected to quote
both the full-split F1 and the clean-subset F1, so that memorisation
cannot be confused with generalisation.

\section{Availability and Reproduction}
\label{sec:availability}

The artefact is archived on Zenodo~\cite{javavulbench2026}.
Data are released under CC-BY-4.0 and code under the MIT licence. The archive is
self-contained: it ships \texttt{data/raw/}, processed splits,
prediction files, result tables, and fine-tuned checkpoints, so reviewers
can reproduce the reported results without GitHub or NVD API access.

Reproduction is driven by \texttt{scripts/reproduce.py}. The pipeline
has five resumable stages: \texttt{data} extracts methods, applies noise
filters, and generates the shipped splits; \texttt{models} downloads
external weights when needed; \texttt{audit} runs the pre-training
contamination audit; \texttt{eval} fine-tunes the encoder baselines and
runs the API-LLM probe; and \texttt{paper} aggregates prediction files
into the CSV tables used in this paper. Stages can be selected with
\texttt{-{}-only} or skipped with \texttt{-{}-skip}; partial runs reuse
on-disk prediction files rather than recomputing them.

On our reference machine, Windows~11, Python~3.12, CUDA~12.8, an
RTX 4060/5080-class GPU with 8-12\,GB VRAM, and 16\,GB RAM, the cached
pipeline completes in about 30-45 minutes when the shipped mining
artifacts are reused. Building the
benchmark with \texttt{-{}-stats} takes about 5 seconds from cached
artifacts, while the pre-training contamination audit takes 30--90
seconds. Fine-tuning \texttt{unixcoder-base} for two epochs on 3k samples
takes 3--5 minutes on GPU; the fallback GPT-4.1-mini 20-sample API probe
takes 20--40 seconds; and the same fine-tuning run on CPU takes
45--75 minutes.

A full rebuild takes about 8-12 hours. The dominant cost is
the mining stage, CVE $\rightarrow$ commit $\rightarrow$ source, which
takes roughly 6-10 hours. Benchmark construction then takes 10-20
minutes on CPU, while the contamination audit, fine-tuning step, API
fallback, and csv generation have the same expected runtimes as
in the cached setting.

Docker pins Python~3.12 and CUDA~12.4. API-based runs require
\texttt{OPENROUTER\_KEY}; mining from scratch requires
\texttt{GITHUB\_TOKEN}. The Zenodo snapshot can regenerate the reported
tables without either key.
\section{Positioning and Limitations}
\label{sec:positioning}

\begin{table}[t]
\caption{Java vulnerability benchmarks. M = method/function-level
labels shipped directly; N = explicit non-vulnerable samples;
L = line-level vulnerable-line labels; S = $\geq 2$ realistic split
strategies shipped; H = unified multi-backend evaluation harness;
A = per-model pre-training contamination audit. \ding{51} = yes,
\ding{55} = no, $\sim$ = partial / derivable from shipped diffs but not
shipped directly.}
\label{tab:related}
\small
\setlength{\tabcolsep}{4pt}
\begin{tabular}{lccccccc}
\toprule
Benchmark                    & Java-only & M & N & L & S & H & A \\
\midrule
Vul4J~\cite{vul4j2022}        & \ding{51} & $\sim$   & \ding{55} & $\sim$   & \ding{55} & \ding{55} & \ding{55} \\
VJBench~\cite{vjbench2023}    & \ding{51} & $\sim$   & \ding{55} & $\sim$   & \ding{55} & \ding{55} & \ding{55} \\
CWE-Bench-Java~\cite{cwebenchjava2024}
                              & \ding{51} & \ding{51} & \ding{55} & \ding{55} & \ding{55} & \ding{55} & \ding{55} \\
CVEfixes~\cite{cvefixes2021}  & \ding{55} & $\sim$   & \ding{55} & $\sim$   & \ding{55} & \ding{55} & \ding{55} \\
\textbf{\textsc{JavaVulBench}} (this work)
                              & \ding{51} & \ding{51} & \ding{51} & \ding{51} & \ding{51} & \ding{51} & \ding{51} \\
\bottomrule
\end{tabular}
\end{table}

\textbf{Positioning.} Table~\ref{tab:related} compares
\textsc{JavaVulBench} with the closest Java vulnerability benchmarks.
Vul4J~\cite{vul4j2022} and VJBench~\cite{vjbench2023} are
repair-oriented corpora with reproducible Java CVEs and PoC tests,
rather than method-level classification datasets with explicit
negatives. CWE-Bench-Java~\cite{cwebenchjava2024} provides
function-level Java CVEs, but lacks paired non-vulnerable samples,
line labels, and shipped realistic splits; CVEfixes~\cite{cvefixes2021}
is multilingual and commit/file-level. \textsc{JavaVulBench} adds the
combination of method+line labels, stratified negatives, five realistic
splits, a unified multi-backend harness, and per-model contamination
auditing for enterprise Java vulnerabilities such as deserialization,
SSRF, path traversal, injection, and authentication/authorisation flaws.

\textbf{Limitations.} Line labels approximate developer-edited fix
lines and may miss semantically vulnerable statements. The
\emph{deduplicated} split is currently equivalent to \emph{random} at
Jaccard 0.8; stronger cross-project duplicate detection is left to the
datasheet. API-LLM results use a deterministic 200-sample per-split
probe for cost reasons, while the fine-tuned baselines in
Table~\ref{tab:rq2_full} remain evaluated on the full test set.

\bibliographystyle{ACM-Reference-Format}
\bibliography{refs}

\end{document}